\documentclass[preprint,showpacs,preprintnumbers,amsmath,amssymb,aps,pre,
floatfix]{revtex4}

\usepackage{graphicx}
\usepackage{dcolumn}
\usepackage{bm}
\def\lap{\hbox{~{\lower -2.5pt\hbox{$<$}}\hskip -8pt\raise -3.5pt\hbox{$\sim$}}}
\def\gap{\hbox{~{\lower -2.5pt\hbox{$>$}}\hskip -8pt\raise -3.5pt\hbox{$\sim$}}}
\def\apg{\hbox{{\raise -2.5pt\hbox{$>$}}\hskip -8pt\lower -2.5pt\hbox{$\sim$}}}
\def\apl{\hbox{{\raise -2.5pt\hbox{$<$}}\hskip -8pt\lower -2.5pt\hbox{$\sim$}}}

\begin{document}

\preprint{LA-UR-06-8651}

\title{Vibrational Contribution to Density and Current Autocorrelations in a 
Monatomic Liquid}

\author{Eric D. Chisolm}
\author{Giulia De Lorenzi-Venneri}
\author{Duane C. Wallace}
\affiliation{Theoretical Division, Los Alamos National Laboratory, 
Los Alamos, New Mexico 87545}

\date{\today}

\begin{abstract}
We consider for a monatomic liquid the density and current
autocorrelation functions from the point of view of the
Vibration-Transit (V-T) theory of liquid dynamics. We also consider
their Fourier transforms, one of which is measured by X-ray and
neutron scattering.  In this description, the motion of atoms in the
liquid is divided into vibrations in a single characteristic potential
valley, called a \emph{random} valley, and nearly-instantaneous
transitions called \emph{transits} between valleys.  The theory
proposes a Hamiltonian for the vibrational motion, to be corrected to
take transits into account; this Hamiltonian is used to calculate the
autocorrelation functions, giving what we call their vibrational
contributions.  We discuss the multimode expansions of the
autocorrelation functions, which provide a physically helpful picture
of the decay of fluctuations in terms of $n$-mode scattering
processes; we also note that the calculation and Fourier transform of
the multimode series are numerically problematic, as successive terms
require larger sums and carry higher powers of the temperature, which
is a concern for the liquid whose temperature is bounded from below by
melt.  We suggest that these problems are avoided by directly
computing the autocorrelation functions, for which we provide
straightforward formulas, and Fourier transforming them numerically.
\end{abstract}

\maketitle

\section{Introduction}
\label{intro}

V-T theory is being developed specifically to describe atomic motion
in real monatomic liquids. This theory adopts a point of view
different from, and in a certain way complementary to, the more
traditional formulations. The theory most successful for equilibrium
thermodynamic properties is based on the interatomic pair potential
and the pair distribution function \cite{ASSSP33,Marchbook}. This
theory can be made exact by including $n$-body potentials for $n\ge3$.
On the other hand, the most widely studied formulation of
nonequilibrium properties, as expressed through time correlation
functions, is generalized hydrodynamics \cite{BYbook,HMCDbook,BZbook}. 
The memory function analysis is formally exact, but has not yet proved
amenable to first-principles evaluation. V-T theory is based on an
approximate but tractable zeroth order Hamiltonian, which in principle
can be systematically improved to include corrections beyond zeroth
order. It complements traditional theory in that both equilibrium and
nonequilibrium properties can be calculated from the same zeroth order
theory, without adjustable parameters \cite{VT5}.

V-T theory is based on a symmetry classification of the potential
energy valleys among which the liquid atoms move \cite{VT5,VT8,VT9}. 
Specifically, the amorphous (noncrystalline) valleys are classified as
(a) symmetric valleys, with e.g.\ microcrystalline structures, which
have significant variations in their potential energy and vibrational
frequencies but are relatively few in number, and (b) random valleys,
with approximately random close-packed structures, which all have the
same potential energy and vibrational frequencies in the thermodynamic
limit, and which are of overwhelming numerical superiority.  As a
zeroth order model for the potential energy surface, we neglect the
symmetric valleys, and construct an ``extended random valley'' as the
harmonic extension to infinity of one (any) random valley. The
partition function is then the extended random valley partition
function, times the number of such valleys. This is a universal
number, calibrated from the constant density melting entropy of the
elements \cite{VT3}. The zeroth order theory gives a good account of
the experimental thermodynamic properties of the elemental liquids at
melt, indicating that the corrections are small at melt \cite{VT5}.

In a real monatomic liquid the motion consists of two parts:
vibrational motion in a random valley, and transit motion, wherein the
system crosses valley-valley intersections. In applying V-T theory to
a physical process, we evaluate the vibrational contribution in zeroth
order, exactly and without adjustable parameters, and then model the
effect of transits. Thus in the liquid partition function the major
contribution is vibrational, while the role of transits is to allow
the system to access the multitude of random valleys, and hence to
possess the extra entropy associated with melting.  In the dynamic
response function we have found that both the location and area of the
Brillouin peak are given by the vibrational contribution, while
transits are responsible only for a partial broadening of this
peak. Hence an a priori calculation of the vibrational contribution to
the Brillouin peak dispersion curve for pseudopotential sodium agrees
within experimental error with the measured results of inelastic x-ray
scattering for real liquid sodium \cite{ARXIV05}. In addition, the
vibrational contribution can be modified to account for transits, and
this model provides an excellent fit to the entire Brillouin peak
\cite{DWJCP123}.

The purpose of the present paper is to study the vibrational
contributions to density and current correlations in a monatomic
liquid. The density autocorrelation function $F(q,t)$ and its Fourier
transform $S(q,\omega)$ are the basis for inelastic neutron scattering
theory. This theory was originally developed for crystals, and
expansions of $S(q,\omega)$ in powers of nuclear displacements were
analyzed in detail by Maradudin and Fein \cite{MFPR128}, Ambegaokar,
Conway, and Baym \cite{ACBLD}, Cowley \cite{CAP12}, Lovesey
\cite{Lovbook}, and Glyde \cite{Glybook}.  A comparison of different
expansions may be found in Glyde \cite{Glybook}.  (Physically
motivated approximations for the glassy state are discussed by
Egelstaff \cite{Egelstaff}.)  Liquid theory has important similarities
to, and differences from, the crystal theory. First, in liquid and
crystal alike, the multimode expansion converges more slowly as the
momentum transfer $q$ increases, and as the temperature $T$
increases. While the range of $q$ is approximately the same for
inelastic scattering from a given material in crystal and liquid form,
the temperature is quite the opposite, being below $T_{m}$ for the
crystal and above $T_{m}$ for the liquid.  Second, vibrational
dynamics is more complicated for a random valley, because every site
is inequivalent, as opposed to inequivalence of only the crystal sites
in one unit cell. As a result, the dynamical matrix is $3N\times3N$
for a random valley, but is block diagonal in $N$ $3\times3$ matrices
for a primitive lattice crystal. We will show how to overcome
convergence problems resulting from the high temperatures of the
liquid, and how to address computability problems resulting from the
large dynamical matrix. On the other hand, we will not address
anharmonic corrections here, as it is first necessary to formulate and
test the harmonic theory. Further, since the atomic motion is
classical in most monatomic liquids, we work with classical
statistical mechanics, which allows us to freely commute atomic
positions and momenta.

The longitudinal and transverse current autocorrelation functions
$C_{L,T}(q,t)$ are discussed by March \cite{Marchbook}, Hansen and
McDonald \cite{HMCDbook}, and Balucani and Zoppi \cite{BZbook}. The
relation of these functions via Green-Kubo theory to the longitudinal
and transverse viscosities is described, as well as their treatment
within the framework of generalized hydrodynamics. We notice that
$C_{L}(q,\omega)=(\omega^{2}/q^{2}) S(q,\omega)$, while
$C_{T}(q,\omega)$ is an independent function \cite{HMCDbook,BZbook}.
For the liquid alkali metals, dispersion curves associated with
$C_{L}(q,\omega)$ have been attributed to sound-like collective modes
of the atomic motion (\cite{BZbook}, Fig.~6.6). These authors also
argue for the existence of distinct fast and slow decay mechanisms in
$C_{L}(q,\omega)$ for monatomic liquids (\cite{BZbook}, Sec.~6.2).  In
V-T theory, these would result from vibrations and transits
respectively.  Dispersion curves associated with $C_T(q,\omega)$ have
also been studied \cite{JMCD80}. For the transverse current
correlation function in liquid rubidium, it has been shown that the
usual phenomenological relaxation time approximations for the memory
functions are not capable of fitting the MD data \cite{BVG87}.

Currently there is no theory, short of computer simulations, that will
account for these dispersion curves without adjustable
parameters. Since the vibrational contribution alone does this for the
Brillouin peak dispersion curve \cite{ARXIV05}, it is possible that it
will do so as well for the dispersion curves associated with
$C_{L}(q,\omega)$ and $C_{T}(q,\omega)$. This possibility motivates
the present work.  In addition, the computational problems mentioned
above for $S(q,\omega)$ need to be addressed also for
$C_{L,T}(q,\omega)$; these are the poor convergence of the multimode
expansions at high temperatures, and the difficulty of evaluating
nested $3N$-fold sums for moderately large $N$.

In Sec.~\ref{vib} we briefly summarize the vibrational normal mode theory for
an extended random valley, and obtain the pair correlation functions
involving atomic positions and velocities. In Sec.~\ref{Sqw} we write the
formulas for vibrational contributions to the dynamic response
functions, $F(q,t)$ and $S(q,\omega)$; we compare these with the
formulas derived by Carpenter and Pelizzari \cite{CPPRB12}; and we
show how to circumvent practical limitations posed by convergence and
computability problems. The vibrational contributions to
$C_{L,T}(q,t)$ are derived in closed form in Sec.~\ref{Cqw}, and are shown to
consist of decoupled velocity-position terms and velocity-position
interference terms. The one- and two-mode contributions to
$C_{L,T}(q,\omega)$ are derived, and accurate but practical evaluation
procedures are described for all the current correlation functions.  A
brief summary is presented in Sec.~\ref{summ}.

\section{Random Valley Vibrational Modes}
\label{vib}

This topic has been discussed previously \cite{VT13}, so a brief
outline will suffice here and will establish our notation. Our
mechanical system consists of $N$ atoms in a cubical box, with the
atomic motion governed by periodic boundary conditions at the box
surface. The atomic positions and momenta at time $t$ are respectively
$\bm{r}_{K}(t)$ and $\bm{p}_{K}(t)$ for $K=1,\dots, N$. For motion in
a single random valley we follow the lattice dynamics tradition of
Born and Huang \cite{BornHuangbook} and write
\begin{equation}\label{eq1}
\bm{r}_{K}(t) = \bm{R}_{K} + \bm{u}_{K}(t),
\end{equation}
where $\bm{R}_{K}$ is the equilibrium position and $\bm{u}_{K}(t)$ is
the displacement. The system potential is $\Phi(\{\bm{r}_{K}\})$ and
the potential at the equilibrium structure is
$\Phi_{0}=\Phi(\{\bm{R}_{K}\})$.  The extended random valley is by
definition harmonic, and has the Hamiltonian
\begin{equation}\label{eq2}
{\mathcal H}_{vib} = \Phi_{0} + \sum_{Ki} \frac{p_{Ki}^{2}}{2M} + 
\frac{1}{2} \sum_{KL} \sum_{ij} \Phi_{Ki,Lj} u_{Ki} u_{Lj}, 
\end{equation}
where $i,j$ represent Cartesian directions and $M$ is the atomic
mass. The potential energy coefficients $\Phi_{Ki,Lj}$ form a real
symmetric $3N\times3N$ matrix, which is the dynamical matrix
\cite{BornHuangbook} for the random valley. The vibrational normal
modes have coordinates and momenta respectively $q_{\lambda}(t)$ and
$p_{\lambda}(t)$, for $\lambda=1,\dots,3N$.  The transformation from
displacements is defined by
\begin{eqnarray}\label{eq3}
q_{\lambda}& =&\sum_{Ki}w_{Ki,\lambda}\;u_{Ki}, \nonumber \\ 
p_{\lambda}&=&\sum_{Ki}w_{Ki,\lambda}\;p_{Ki},
\end{eqnarray}
where the eigenvector components $w_{Ki,\lambda}$ are real and
diagonalize the dynamical matrix. It follows that
\begin{equation}\label{eq4}
{\mathcal H}_{vib}=\Phi_{0}+ \sum_{\lambda} \left[\frac{p_{\lambda}^{2}}{2M} + 
\frac{1}{2} M\omega_{\lambda}^{2}q_{\lambda}^{2}\right],
\end{equation}
where $\omega_{\lambda}$ are the mode frequencies, from which it also
follows that
\begin{eqnarray}
q_\lambda(t) & = & q_\lambda(0) \cos \omega_\lambda t + \frac{p_\lambda(0)}
                  {M \omega_\lambda} \sin \omega_\lambda t \nonumber \\
p_\lambda(t) & = & p_\lambda(0) \cos \omega_\lambda t - M \omega_\lambda 
                  q_\lambda(0) \sin \omega_\lambda t.
\label{eq4.5}
\end{eqnarray} 
The eigenvector components also satisfy orthonormality and
completeness relations \cite{VT13}, respectively given by
\begin{equation}\label{eq5}
\sum_{Ki} w_{Ki,\lambda} w_{Ki,\lambda'} = \delta_{\lambda \lambda'},
\end{equation}
\begin{equation}\label{eq6}
\sum_{\lambda} w_{Ki,\lambda} w_{Lj,\lambda} = \delta_{KL}\delta_{ij}.
\end{equation}

The basic ingredients of vibrational time correlation functions are
the mode-mode functions such as $\left<q_{\lambda}(t)q_{\lambda'}(0)\right>$. 
These can be calculated from Eqs.~(\ref{eq4}) and (\ref{eq4.5}), in
classical or quantum mechanics, where the classical limit of the
quantum result is the leading term in the expansion for
$kT/\hbar\omega_{\lambda}>1$. For $-\infty<t<\infty$, the classical 
expressions are
\begin{equation}\label{eq7}
\left< q_{\lambda}(t) q_{\lambda'}(0) \right>= \frac{kT}{M\omega_{\lambda}^{2}} 
\delta_{\lambda\lambda'}\cos\omega_{\lambda}t,\\
\end{equation}
\begin{equation}\label{eq8}
\left< p_{\lambda}(t) \;p_{\lambda'}(0) \right>= MkT \delta_{\lambda\lambda'}
\cos\omega_{\lambda}t,\\
\end{equation}
\begin{equation}\label{eq9}
\left< p_{\lambda}(t) q_{\lambda'}(0) \right>= -\frac{kT}{\omega_{\lambda}} 
\delta_{\lambda\lambda'}\sin\omega_{\lambda}t.
\end{equation}
Correlation functions involving the atomic positions and momenta
follow from these and the normal mode transformation,
Eqs.~(\ref{eq3}). It will be convenient to use velocity
$\dot{\bm{u}}_{K}$ instead of $\bm{p}_{K}=M\dot{\bm{u}}_{K}$, and to
eliminate the Cartesian components in favor of vector dot
products. Then with $\bm{a}$ and $\bm{b}$ arbitrary vectors,
\begin{equation}\label{eq10}
\left<\bm{a}\cdot\dot{\bm{u}}_{K}(t)\;\bm{b}\cdot\dot{\bm{u}}_{L}(0)\right> = 
 \frac{kT}{M}\sum_{\lambda} \bm{a}\cdot \bm{w}_{K\lambda}\; \bm{b}\cdot 
 \bm{w}_{L\lambda} \cos\omega_{\lambda}t,
\end{equation}
\begin{equation}\label{eq11}
\left<\bm{a}\cdot\dot{\bm{u}}_{K}(t)\;\bm{b}\cdot\bm{u}_{L}(0)\right> = 
- \frac{kT}{M}\sum_{\lambda} \bm{a}\cdot \bm{w}_{K\lambda}\; \bm{b}\cdot 
 \bm{w}_{L\lambda} \frac{\sin\omega_{\lambda}t}{\omega_{\lambda}},
\end{equation}
\begin{equation}\label{eq12}
\left<\bm{a}\cdot\bm{u}_{K}(t)\;\bm{b}\cdot\bm{u}_{L}(0)\right> = 
 \frac{kT}{M}\sum_{\lambda} \bm{a}\cdot \bm{w}_{K\lambda}\; \bm{b}\cdot 
 \bm{w}_{L\lambda} \frac{\cos\omega_{\lambda}t}{\omega_{\lambda}^{2}}.
\end{equation}
The fluctuation results are obtained by setting $t=0$. Notice
Eq.~(\ref{eq11}), the velocity-displacement correlation function, is
odd in $t$, and the fluctuation vanishes. Eq.~(\ref{eq11}) satisfies
\begin{equation}\label{eq13}
\left<\bm{a}\cdot\bm{u}_{K}(t)\;\bm{b}\cdot\dot{\bm{u}}_{L}(0)\right> = 
-\left<\bm{a}\cdot\dot{\bm{u}}_{K}(t)\;\bm{b}\cdot\bm{u}_{L}(0)\right>,
\end{equation}
which is valid for any motion as a consequence of time translation
invariance.  Eq.~(\ref{eq11}) also satisfies
\begin{equation}\label{eq13.5}
\left<\bm{a}\cdot\bm{u}_{K}(0)\;\bm{b}\cdot\dot{\bm{u}}_{L}(t)\right> = 
\left<\bm{a}\cdot\dot{\bm{u}}_{K}(t)\;\bm{b}\cdot\bm{u}_{L}(0)\right>,
\end{equation}
but this relation follows from the specific form of Eq.~(\ref{eq11})
and is not true for general motions.  The above equations are valid at
all temperatures for which the atomic motion is classical. Hence these
equations are appropriate for comparison with ordinary (classical) MD
calculations at all $T\ge 0$.

\section{Dynamic Structure Factor}
\label{Sqw}

Because the system motion is constrained by periodic boundary
conditions, Fourier components are defined only at a discrete set of
$\bm{q}$. To reflect the isotropic nature of a real liquid, a
macroscopic average can include an average over the star of $\bm{q}$
for each $|\bm{q}|$, this average being denoted
$\left<\cdots\right>_{\bm{q}^{\ast}}$.  The density autocorrelation
function, also called the van Hove function, is then written
\begin{equation}\label{eq14}
F(q,t) = \frac{1}{N}
 \left < \left< \sum_{K} e^{-i \bm{q}\cdot \bm{r}_{K}(t)} 
 \sum_{L} e^{i \bm{q}\cdot \bm{r}_{L}(0)} \right>\right >_{\bm{q}^{\ast}}.
\end{equation}
Here each sum is a $\bm{q}$-component of the system density, and the
inner average is over the vibrational motion, as in
Eqs.~(\ref{eq7}-\ref{eq13.5}). Notice the $\bm{q}^{\ast}$ average makes
$F(q,t)$ a function only of the magnitude $q=|\bm{q}|$. We have
previously used notation such as $F_{vib}(q,t)$ and
$\left<\cdots\right>_{vib}$, but here the vibrational subscript will
be omitted.  Then inserting Eq.~(\ref{eq1}) for $\bm{r}_{K}(t)$,
\begin{equation}\label{eq15}
F(q,t) = \frac{1}{N} \left< \sum_{KL} e^{-i \bm{q}\cdot \bm{R}_{KL}} 
 \left < e^{-i \bm{q} \cdot (\bm{u}_{K}(t) - \bm{u}_{L}(0))}\right> 
 \right >_{\bm{q}^{\ast}},
\end{equation}
where $\bm{R}_{KL}=\bm{R}_{K}-\bm{R}_{L}$. The vibrational average is
evaluated by Bloch's identity (e.g. \cite{Lovbook}), which is derived
via the normal mode transformation, to write
\begin{equation}\label{eq16}
F(q,t) =	\frac{1}{N} \left<  \sum_{KL} e^{-i \bm{q}\cdot \bm{R}_{KL}} 	
   e^{-\frac{1}{2} \left <\left [ \bm{q}\cdot(\bm{u}_{K}(t)-\bm{u}_{L}(0)) \right ]^{2} 
    \right>} \right >_{\bm{q}^{\ast}}.
\end{equation}
Finally we separate the fluctuation terms to find
\begin{equation}\label{eq17}
F(q,t)=	\frac{1}{N} \left<  \sum_{KL} e^{-i \bm{q}\cdot \bm{R}_{KL}} 	
  e^{-W_{K}(\bm{q})} e^{-W_{L}(\bm{q})}  e^{\left < \bm{q} \cdot \bm{u}_{K}(t)\;\bm{q} 
   \cdot \bm{u}_{L}(0)\right >}\right >_{\bm{q}^{\ast}}
\end{equation}
where $e^{-W_{K}(\bm{q})}$ is the Debye-Waller factor for atom $K$,
which can be expressed with the aid of Eq.~(\ref{eq12}) as
\begin{equation}\label{eq18}
W_{K}(\bm{q})=\frac{kT}{2M}\sum_{\lambda}\frac{(\bm{q}\cdot\bm{w}_{K\lambda})^{2}}
             {\omega_{\lambda}^{2}}.
\end{equation}
The function in the exponential may also be evaluated using
Eq.~(\ref{eq12}).

The time dependence of $F(q,t)$ is contained in the time correlation
functions \linebreak
$\left<\bm{q}\cdot\bm{u}_{K}(t)\;\bm{q}\cdot\bm{u}_{L}(0)\right>$ in
Eq.~(\ref{eq17}). The multimode expansion is obtained by expanding in
powers of these functions, according to
\begin{equation}\label{eq19}
 e^{\left < \bm{q} \cdot \bm{u}_{K}(t)\;\bm{q} \cdot \bm{u}_{L}(0)\right >}=1 + 
 \left < \bm{q} \cdot \bm{u}_{K}(t)\;\bm{q} \cdot \bm{u}_{L}(0)\right >+
 \frac{1}{2}\left < \bm{q} \cdot \bm{u}_{K}(t)\;\bm{q} \cdot \bm{u}_{L}(0)
 \right >^{2} + \cdots .
\end{equation}
This is a useful procedure, because each term can be Fourier
transformed analytically, to give an explicit picture of the
scattering process in each order.  For a random valley,
$\left<\bm{q}\cdot\bm{u}_{K}(t)\;\bm{q}\cdot\bm{u}_{L}(0)\right>
\rightarrow0$ as $t\rightarrow \infty$, so that $F(q,\infty)$ is
positive and is given by
\begin{equation}\label{eq20}
F(q,\infty)=\frac{1}{N} \left< \left| \sum_{K}  e^{-i \bm{q}\cdot \bm{R}_{K}}\;
e^{-W_{K}(\bm{q})}\right |^{2}\right>_{\bm{q}^{\ast}}.
\end{equation}
Then the series for $S(q,\omega)$ is
\begin{equation}\label{eq21}
S(q,\omega) = F(q,\infty)\delta(\omega)+S^{(1)}(q,\omega) + S^{(2)}(q,\omega)+
\cdots .
\end{equation}
The leading term expresses elastic scattering. $S^{(1)}(q,\omega)$
arises from the linear term in Eq.~(\ref{eq19}), and with
Eq.~(\ref{eq12}) the expression can be organized into
\begin{equation}\label{eq22}
S^{(1)}(q,\omega)=\frac{kT}{M}\frac{1}{N}\sum_{\lambda}\left<|f_{\lambda}(q)|^2
\right>_{\bm{q}^{\ast}} \frac{1}{2}\Big[\delta(\omega+\omega_{\lambda})+
\delta(\omega-\omega_{\lambda})\Big], 
\end{equation}
where
\begin{equation}\label{eq23}
f_{\lambda}(q)=\frac{1}{\omega_{\lambda}}\sum_{K}  e^{-i \bm{q}\cdot \bm{R}_{K}}
\;e^{-W_{K}(\bm{q})}\;\bm{q} \cdot \bm{w}_{K\lambda}.
\end{equation}
(Note that this $f_\lambda(q)$ differs from our definition in
\cite{ARXIV05,DWJCP123}, although $S^{(1)}(q,\omega)$ is the same.)
Hence $S^{(1)}(q,\omega)$ is the sum over all vibrational modes of
single-mode scattering at momentum transfer $\hbar q$, in which energy
$\hbar\omega_{\lambda}$ is either created in mode $\lambda$ (the
$\delta(\omega+\omega_{\lambda})$ term), or is annihilated in mode
$\lambda$ (the $\delta(\omega-\omega_{\lambda})$ term).

In a similar way, $S^{(2)}(q,\omega)$ can be written
\begin{equation}\label{eq24}
S^{(2)}(q,\omega)=\left(\frac{kT}{M}\right)^{2}\frac{1}{2N}\sum_{\lambda\lambda'}
\left< |f_{\lambda\lambda'}(q)|^2\right>_{\bm{q}^{\ast}} 
\left<\delta(\omega\pm\omega_{\lambda}\pm\omega_{\lambda'})\right>_{S},
\end{equation}
where
\begin{equation}\label{eq25}
f_{\lambda\lambda'}(q)=\frac{1}{\omega_{\lambda}\omega_{\lambda'}}
\sum_{K} e^{-i \bm{q}\cdot \bm{R}_{K}}\;e^{-W_{K}(\bm{q})}\;\bm{q} \cdot 
\bm{w}_{K\lambda}\;\bm{q} \cdot \bm{w}_{K\lambda'},
\end{equation}
and $\left<\delta(\omega\pm \omega_{\lambda}\pm\omega_{\lambda'})
\right>_{S}$ is the symmetric average of four $\delta$-functions,
\begin{eqnarray}\label{eq26}
\left<\delta(\omega\pm\omega_{\lambda}\pm\omega_{\lambda'})\right>_{S}  &= & 
 \frac{1}{4} \Big[\delta(\omega+\omega_{\lambda}+\omega_{\lambda'}) + 
 \delta(\omega-\omega_{\lambda}-\omega_{\lambda'})  \nonumber\\
& & + \delta(\omega+\omega_{\lambda}-\omega_{\lambda'})+\delta(\omega-
 \omega_{\lambda}+\omega_{\lambda'})\Big].
\end{eqnarray}
$S^{(2)}$ is therefore the sum over all pairs of vibrational modes of
the cross section for scattering at momentum transfer $\hbar q$, in
which two excitations are created (the $\delta(\omega+\omega_{\lambda}+
\omega_{\lambda'})$ term) or annihilated (the $\delta(\omega-
\omega_{\lambda}-\omega_{\lambda'})$ term), or in which one excitation
is created and another is annihilated (the $\delta(\omega\pm(\omega_{\lambda} 
-\omega_{\lambda'}))$ terms).

In the multimode expansion of $S(q,\omega)$ we see from
Eqs.~(\ref{eq22}) and (\ref{eq24}) that $S^{(n)}(q,\omega)$ is
proportional to $T^{n}$. Therefore any finite termination of the
series diverges with increasing $T$ even though $S(q,\omega)$ does
not. We also see from the above equations that $S^{(n)}(q,\omega)$
requires evaluation of $n+1$ nested sums, each over atoms or over
normal modes, so that each sum contains of order of $N$ terms. To
characterize the magnitude of the computation, we consider a system of
$N=1000$ atoms, for which the data $\{\bm{R}_{K};\omega_{\lambda};
\bm{w}_{K\lambda}\}$ are available. Then single, double, and triple
sums are readily computed, while a quadruple sum challenges current
computer facilities. On the other hand, problems related to both
convergence and computability of the multimode expansion are
eliminated by writing
\begin{equation}
F(q,t) = \left[F(q,t)-F(q,\infty)\right] + F(q,\infty) 
\end{equation}
where $F(q,\infty)$ is given in Eq.~(\ref{eq20}); the first term can
be Fourier transformed numerically, and the transform of the second
term is just $F(q,\infty)\delta(\omega)$ (see Eq.~(\ref{eq21})).
Since the time correlation functions in Eq.~(\ref{eq17}) are
proportional to $T$, $F(q,t)$ is well behaved as a function of
$T$. Further, $F(q,t)$ can be computed exactly as a triple sum, as can
any term in the multimode expansion of $F(q,t)$.

The quantum expressions for $S^{(1)}(q,\omega)$ and
$S^{(2)}(q,\omega)$ for a polyatomic amorphous solid were derived by
Carpenter and Pelizzari \cite{CPPRB12}. When their results are applied
to a monatomic system and the classical limit is taken, our
Eqs.~(\ref{eq22}-\ref{eq26}) are obtained with the following minor
differences. Our eigenvectors are real (not complex) and are
normalized to 1 (not $N$); our $\lambda'=\lambda$ term in
Eq.~(\ref{eq24}) contains the two inelastic terms of Carpenter and
Pelizzari, plus two additional elastic terms in $\delta(\omega)$; and
by writing the $\sum_{KL}$ in Eq.~(\ref{eq17}) as
$|\sum_{K}\cdots|^{2}$ in Eqs.~(\ref{eq22}-\ref{eq25}), our
multimode terms have one lower order of nested sums than those of
Carpenter and Pelizzari.

\section{Current Autocorrelation Functions}
\label{Cqw}

Fourier components of the particle current are denoted
$\bm{j}(\bm{q},t)$, where \cite{HMCDbook,BZbook,Marchbook}
\begin{equation}\label{eq27}
\bm{j}(\bm{q},t) = \sum_{K}\dot{\bm{r}}_{K}(t)\; e^{-i \bm{q}\cdot \bm{r}_{K}(t)}.
\end{equation}
Let us define for each $\bm{q}$ the orthogonal set of unit vectors
$\hat{\bm{q}},\hat{\bm{s}}_{1}$ and $\hat{\bm{s}}_{2}$.  Then the
longitudinal and transverse currents are
\begin{eqnarray}\label{eq28}
\bm{j}_{L}(\bm{q},t)& = & \hat{\bm{q}} \hat{\bm{q}}\cdot\bm{j}(\bm{q},t), 
\nonumber \\
\bm{j}_{T}(\bm{q},t)& = &\hat{\bm{s}}_{1}\hat{\bm{s}}_{1}\cdot\bm{j}(\bm{q},t) 
+ \hat{\bm{s}}_{2} \hat{\bm{s}}_{2}\cdot\bm{j}(\bm{q},t).
\end{eqnarray}
The corresponding current autocorrelation functions are defined by
\cite{BZbook}
\begin{eqnarray}\label{eq29}
C_{L}(q,t) & = & \frac {1}{N} \left < \left <\bm{j}_{L} (\bm{q},t) \cdot 
\bm{j}_{L}^{\ast} (-\bm{q},0) \right > \right >_{\bm{q}^{\ast}}, \nonumber \\
C_{T}(q,t) & = & \frac {1}{2N} \left <\left < \bm{j}_{T} (\bm{q},t)\cdot 
\bm{j}_{T}^{\ast} (-\bm{q},0) \right > \right >_{\bm{q}^{\ast}}.
\end{eqnarray}
In these defining equations the inner averages are over the complete 
atomic motion for a macroscopic system. In what follows we shall limit 
ourselves to the vibrational motion in a finite system with periodic 
boundary conditions.

Let $\hat{\bm{a}}$ represent any one of $\hat{\bm{q}}$, $\hat{\bm{s_{1}}}$,
$\hat{\bm{s_{2}}}$, and define the autocorrelation function
\begin{equation}\label{eq30}
D(\hat{\bm{a}},q,t) = \frac{1}{N}\left <\left < \sum_{K}\hat{\bm{a}}\cdot
\dot{\bm{r}}_{K}(t) \; e^{-i \bm{q}\cdot \bm{r}_{K}(t)}
\sum_{L}\hat{\bm{a}}\cdot\dot{\bm{r}}_{L}(0) \; e^{i \bm{q}\cdot \bm{r}_{L}(0)}
\right > \right >_{\bm{q}^{\ast}},
\end{equation}
where the inner average is over vibrations in an extended random
valley.  Then both $C_{L}$ and $C_{T}$ are represented by $D$, since
\begin{eqnarray}\label{eq31}
C_{L}(q,t) &=& D(\hat{\bm{q}},q,t) \nonumber \\
C_{T}(q,t)& = &\frac{1}{2} \left [ D(\hat{\bm{s}}_{1},q,t) +  
D(\hat{\bm{s}}_{2},q,t) \right] =  D(\hat{\bm{s}}_{1},q,t).
\end{eqnarray}
The last equality reflects the isotropic symmetry of a real liquid.

With Eq.~(\ref{eq1}) for $\bm{r}_{K}(t)$, Eq.~(\ref{eq30}) becomes
\begin{equation}\label{eq32}
D(\hat{\bm{a}},q,t) = \frac{1}{N}\left < \sum_{KL}e^{-i \bm{q}\cdot \bm{R}_{KL}}
\left<\left(\hat{\bm{a}}\cdot\dot{\bm{u}}_{K}(t)\;\hat{\bm{a}}\cdot
\dot{\bm{u}}_{L}(0)\right) \;e^{-i \bm{q}\cdot(\bm{u}_{K}(t)-\bm{u}_{L}(0))}\right> 
\right>_{\bm{q}_{\ast}}. 
\end{equation}
The average is quadratic in velocities, and contains terms of all
positive powers in displacements. The odd powers are imaginary, and
harmonic averages involving them vanish. Hence the nonzero averages
are quadratic in velocities and of all even powers in displacements.
To evaluate the thermal average in Eq.~(\ref{eq32}), we first use the
normal mode expansion from Eq.~(\ref{eq3}), with the result
\begin{eqnarray}
\lefteqn{\left\langle \hat{\bm{a}}\cdot\dot{\bm{u}}_K(t)\;\hat{\bm{a}}\cdot
      \dot{\bm{u}}_L(0)\;e^{-i \bm{q}\cdot[\bm{u}_K(t)-\bm{u}_L(0)]} 
      \right\rangle} \nonumber \\
& = & \sum_{\lambda, \lambda'} \hat{\bm{a}}\cdot\bm{w}_{K\lambda}\;\hat{\bm{a}}
      \cdot\bm{w}_{L\lambda'} \left \langle \dot{q}_\lambda(t) 
      \dot{q}_{\lambda'}(0) \exp\left[ \sum_\mu -i\bm{q}\cdot\bm{w}_{K\mu} 
      q_\mu(t) + i\bm{q}\cdot\bm{w}_{L\mu} q_\mu(0)\right] \right\rangle
\label{eq32.2}
\end{eqnarray}
where the vectors $\bm{w}_{K\lambda}$ have components
$w_{Ki,\lambda}$.  Eliminating $q_\lambda(t)$ and $\dot{q}_\lambda(t)$
using Eqs.~(\ref{eq4.5}) and replacing $q_\lambda(0)$ and
$p_\lambda(0)$ by $q_\lambda$ and $p_\lambda$ respectively yields
\begin{equation} \label{eq32.3}
\sum_{\lambda, \lambda'} \hat{\bm{a}}\cdot\bm{w}_{K\lambda} \hat{\bm{a}}\cdot
\bm{w}_{L\lambda'} \left\langle \left(\frac{\cos \omega_\lambda t}{M^2} 
p_\lambda p_{\lambda'} - \frac{\omega_\lambda \sin \omega_\lambda t}{M} 
q_\lambda p_{\lambda'}\right)
\exp\left(\sum_\mu A_\mu q_\mu+B_\mu p_\mu\right) \right\rangle
\end{equation}
where the coefficients $A_\mu$ and $B_\mu$ can easily be calculated,
but their exact values are unneeded.  The first term in brackets is
proportional to
\begin{eqnarray} 
\lefteqn{\left\langle  p_\lambda p_{\lambda'}\exp\left(\sum_\mu A_\mu q_\mu+B_\mu 
       p_\mu\right) \right\rangle = \frac{\partial}{\partial B_\lambda} 
       \frac{\partial}{\partial B_{\lambda'}} \left\langle \exp\left(\sum_\mu 
       A_\mu q_\mu+ B_\mu p_\mu\right) \right\rangle} \nonumber \\
 & = & \frac{\partial}{\partial B_\lambda} \frac{\partial}{\partial B_{\lambda'}}
       \exp\left[\frac{1}{2} \left\langle \left( \sum_\mu A_\mu q_\mu+B_\mu p_\mu 
       \right)^2 \right\rangle \right] \nonumber \\
 & = & \left[ \langle p_\lambda p_{\lambda'} \rangle + \left\langle 
       \left(\sum_\nu A_\nu q_\nu+B_\nu p_\nu \right)p_\lambda \right\rangle 
       \left\langle \left(\sum_{\nu'} A_{\nu'} q_{\nu'}+B_{\nu'} p_{\nu'} \right)
       p_{\lambda'} \right\rangle \right] \nonumber \\
 &   & \hspace{0.5in} \times \exp\left[\frac{1}{2} \left\langle \left( \sum_\mu 
       A_\mu q_\mu+B_\mu p_\mu \right)^2 \right\rangle \right] \nonumber \\
 & = & \big[ \langle p_\lambda p_{\lambda'} \rangle - \left\langle \bm{q}\cdot
       [\bm{u}_K(t)-\bm{u}_L(0)]\;p_\lambda \right\rangle 
       \left\langle \bm{q}\cdot[\bm{u}_K(t)-\bm{u}_L(0)]\;p_{\lambda'} 
       \right\rangle \big] \nonumber \\
 &   & \hspace{0.5in} \times \exp\left[-\frac{1}{2} \left\langle \left(\bm{q}
       \cdot[\bm{u}_K(t)-\bm{u}_L(0)]\right)^2 \right\rangle \right].
\label{eq32.4}
\end{eqnarray}
In passing to the second line we used Bloch's identity, 
\begin{equation} \label{eq32.1}
\left\langle \exp\left( \sum_\mu A_\mu q_\mu + B_\mu p_\mu \right) \right\rangle 
= \exp\left[\frac{1}{2} \left\langle \left(\sum_\mu A_\mu q_\mu + B_\mu p_\mu 
\right)^2 \right\rangle \right].
\end{equation}
(Lovesey's proof \cite{Lovbook} is for the quantum mechanical
single-mode case; the generalization to multiple modes is trivial.
The result can also be proved very simply in the classical case by
completing the squares in the integrals.)  In the third line we
relabelled dummy indices, and in the final line we returned the
quantity in the exponential to its original form.  A similar
calculation shows that the other term can be expressed
\begin{eqnarray}
\lefteqn{\left\langle  q_\lambda p_{\lambda'}\exp\left(\sum_\mu A_\mu q_\mu+B_\mu 
        p_\mu\right) \right\rangle = \frac{\partial}{\partial A_\lambda} 
        \frac{\partial}{\partial B_{\lambda'}} \exp\left[\sum_\mu \frac{1}{2}
        \langle (A_\mu q_\mu+B_\mu p_\mu)^2 \rangle\right]} \nonumber \\
 & = & \big[ \langle q_\lambda p_{\lambda'} \rangle - \left\langle \bm{q}\cdot
       [\bm{u}_K(t)-\bm{u}_L(0)]\;q_\lambda \right\rangle 
       \left\langle \bm{q}\cdot[\bm{u}_K(t)-\bm{u}_L(0)]\;p_{\lambda'} 
       \right\rangle \big] \nonumber \\
 &   & \hspace{0.5in} \times \exp\left[-\frac{1}{2} \left\langle \left(\bm{q}
       \cdot[\bm{u}_K(t)-\bm{u}_L(0)]\right)^2 \right\rangle \right].
\label{eq32.5}
\end{eqnarray}
Putting this back together and rearranging, we get
\begin{eqnarray}
\lefteqn{\left\langle \hat{\bm{a}}\cdot\dot{\bm{u}}_K(t)\;\hat{\bm{a}}\cdot
       \dot{\bm{u}}_L(0)\;e^{-i \bm{q}\cdot[\bm{u}_K(t)-\bm{u}_L(0)]} \right\rangle} 
       \nonumber \\
 & = & \Bigg\{ \sum_{\lambda, \lambda'} \hat{\bm{a}}\cdot\bm{w}_{K\lambda}\;
       \hat{\bm{a}}\cdot\bm{w}_{L\lambda'} \bigg[ \left\langle 
       \left(\frac{\cos \omega_\lambda t}{M^2} p_\lambda - 
       \frac{\omega_\lambda \sin \omega_\lambda t}{M} q_\lambda \right) 
       p_{\lambda'} \right\rangle \nonumber \\
 &   & \hspace{0.25in} - \left\langle \left( \frac{\cos \omega_\lambda t}{M^2} 
       p_\lambda - \frac{\omega_\lambda \sin \omega_\lambda t}{M} 
       q_\lambda \right) \bm{q}\cdot[\bm{u}_K(t)-\bm{u}_L(0)] \right\rangle 
       \left\langle \bm{q}\cdot[\bm{u}_K(t)-\bm{u}_L(0)]\;p_{\lambda'}
       \right\rangle \bigg] \Bigg\} \nonumber \\
 &   & \hspace{0.25in} \times \exp\left[-\frac{1}{2} \left\langle 
       \left(\bm{q}\cdot[\bm{u}_K(t)-\bm{u}_L(0)]\right)^2 \right\rangle 
       \right].
\label{eq32.6}
\end{eqnarray}
Next, we change $q_\lambda$ and $p_\lambda$ back to $q_\lambda(0)$ and
$p_\lambda(0)$ and use Eq.~(\ref{eq4.5}) to simplify the
time-dependent factors, reducing the expression to
\begin{eqnarray}
 & & \Bigg\{ \sum_{\lambda, \lambda'} \hat{\bm{a}}\cdot\bm{w}_{K\lambda}\;
     \hat{\bm{a}}\cdot\bm{w}_{L\lambda'} \bigg[ \left\langle 
     \dot{q}_\lambda(t) \dot{q}_{\lambda'}(0) \right\rangle - \left\langle 
     \bm{q}\cdot[\bm{u}_K(t)-\bm{u}_L(0)]\;\dot{q}_\lambda(t) 
     \right\rangle \nonumber \\
 & & \hspace{0.25in} \times \left\langle \bm{q}\cdot[\bm{u}_K(t)-\bm{u}_L(0)]\;
     \dot{q}_{\lambda'}(0) \right\rangle \bigg] \Bigg\} \exp\left[-\frac{1}{2} 
     \left\langle \left(\bm{q}\cdot[\bm{u}_K(t)-\bm{u}_L(0)]\right)^2 
     \right\rangle \right].
\label{eq32.7}
\end{eqnarray}
When this is summed over $\lambda$ and $\lambda'$ using Eq.~(\ref{eq3}), 
the expression becomes
\begin{eqnarray}
 &   & \left[ \left\langle \hat{\bm{a}}\cdot\dot{\bm{u}}_K(t)\;\hat{\bm{a}}
       \cdot\dot{\bm{u}}_L(0) \right\rangle - \left\langle \bm{q}\cdot
       [\bm{u}_K(t)-\bm{u}_L(0)]\;\hat{\bm{a}}\cdot\dot{\bm{u}}_K(t) 
       \right\rangle \left\langle \bm{q}\cdot[\bm{u}_K(t)-\bm{u}_L(0)]\;
       \hat{\bm{a}}\cdot\dot{\bm{u}}_L(0) \right\rangle \right] \nonumber \\
 &   & \hspace{0.25in} \times \exp\left[-\frac{1}{2} \left\langle 
       \left(\bm{q}\cdot[\bm{u}_K(t)-\bm{u}_L(0)]\right)^2 \right\rangle 
       \right].
\label{eqn32.8}
\end{eqnarray}
The second term, involving products of displacements and velocities,
can be simplified somewhat.  Terms such as $\langle
\bm{q}\cdot\bm{u}_K(t)\;\hat{\bm{a}}\cdot\dot{\bm{u}}_K(t) \rangle$
are invariant under time translation, so they can be evaluated at
$t=0$.  In this case, the expression is a linear combination of terms
of the form $\langle q_\lambda p_{\lambda'} \rangle$, all of which
vanish.  Thus the two equal-time terms drop out, so
\begin{eqnarray}
\lefteqn{\left\langle \hat{\bm{a}}\cdot\dot{\bm{u}}_K(t)\;\hat{\bm{a}}\cdot
       \dot{\bm{u}}_L(0)\;e^{-i \bm{q}\cdot[\bm{u}_K(t)-\bm{u}_L(0)]} \right\rangle} 
       \nonumber \\
 & = & [ \langle \hat{\bm{a}}\cdot\dot{\bm{u}}_K(t)\;\hat{\bm{a}}\cdot
       \dot{\bm{u}}_L(0) \rangle + \langle \bm{q}\cdot\bm{u}_L(0)\;
       \hat{\bm{a}}\cdot\dot{\bm{u}}_K(t) \rangle \langle \bm{q}\cdot
       \bm{u}_K(t)\;\hat{\bm{a}}\cdot\dot{\bm{u}}_L(0) \rangle ] \nonumber \\
 &   & \hspace{0.25in} \times \exp\left[-\frac{1}{2} \left\langle 
       \left(\bm{q}\cdot[\bm{u}_K(t)-\bm{u}_L(0)]\right)^2 \right\rangle 
       \right].
\label{eqn32.9}
\end{eqnarray}
The exponential is the same term that appears in Eq.~(\ref{eq16}), so
it may be transformed the same way.  Putting all this together,
Eq.~(\ref{eq32}) becomes
\begin{eqnarray}\label{eq34}
\lefteqn{D(\hat{\bm{a}},q,t) = \frac{1}{N}\Bigg < 
\sum_{KL} e^{-i \bm{q}\cdot \bm{R}_{KL}} e^{-W_{K}(\bm{q})} e^{-W_{L}(\bm{q})}  
e^{\left< \bm{q} \cdot \bm{u}_{K}(t)\;\bm{q} \cdot \bm{u}_{L}(0)\right >}}\nonumber \\
& & \times \;\Big [ \left<\hat{\bm{a}}\cdot\dot{\bm{u}}_{K}(t)\;
\hat{\bm{a}}\cdot\dot{\bm{u}}_{L}(0)\right> +
\left<\hat{\bm{a}}\cdot\dot{\bm{u}}_{K}(t)\;\bm{q}\cdot\bm{u}_{L}(0)\right>
\left<\hat{\bm{a}}\cdot\dot{\bm{u}}_{L}(0)\;\bm{q}\cdot\bm{u}_{K}(t)\right> 
\Big] \Bigg >_{\bm{q}^{\ast}},
\end{eqnarray}
and all of the correlation functions may be evaluated with the aid of
Eqs.~(\ref{eq10}-\ref{eq12}).  Comparison with Eq.~(\ref{eq17})
verifies the well known relation for the longitudinal correlation
function \cite{BZbook},
\begin{equation}\label{eq35}
D(\hat{\bm{q}},q,t) = -\frac{1}{q^{2}}\ddot{F}(q,t).
\end{equation}

Just as with $F(q,t)$, expansion of the time-dependent exponential in
Eq.~(\ref{eq34}) produces a series which can be Fourier transformed
analytically, term by term.  We express this trasformed series as
\begin{equation}\label{eq36}
D(\hat{\bm{a}},q,\omega) = D^{(1)}(\hat{\bm{a}},q,\omega) + 
D^{(2)}(\hat{\bm{a}},q,\omega) + \cdots,
\end{equation}
where the superscript reveals the number of normal modes coupled in
the correlation.  $D^{(1)}$ is the pure velocity term, given by
\begin{equation}\label{eq37}
D^{(1)}(\hat{\bm{a}},q,\omega) = \frac{kT}{M} \frac{1}{N} \sum_{\lambda} 
\left< \left| h_{\lambda}(\hat{\bm{a}},\bm{q})\right| ^{2} \right>_{\bm{q}^{\ast}}
\;\frac{1}{2}\Big[\delta(\omega+\omega_{\lambda})+\delta(\omega-\omega_{\lambda})
\Big],
\end{equation}
where
\begin{equation}\label{eq38}
h_{\lambda}(\hat{\bm{a}},\bm{q}) = \sum_{K} e^{-i \bm{q}\cdot \bm{R}_{K}}
  e^{-W_{K}(\bm{q})}\;\hat{\bm{a}}\cdot\bm{w}_{K\lambda}.
\end{equation}
The similarity with one mode scattering as described by
Eqs.~(\ref{eq22}) and (\ref{eq23}) is apparent. The disappearance here
of the $\omega_{\lambda}^{-1}$ factor in Eq.~(\ref{eq23}) arises from
velocities replacing displacements.

The two-mode contribution takes the form
\begin{eqnarray}\label{eq39}
\lefteqn{D^{(2)}(\hat{\bm{a}},q,\omega) = \left(\frac{kT}{M}\right)^{2} 
\frac{1}{N} \sum_{\lambda \lambda'}\frac{1}{\omega _{\lambda'}^{2}} 
\left<  h_{\lambda \lambda'}(\hat{\bm{a}},\bm{q})\;h_{\lambda \lambda'}^{\ast}
(\hat{\bm{a}},\bm{q})\right>_{\bm{q}^{\ast}}
\left<\delta(\omega\pm\omega_{\lambda}\pm\omega_{\lambda'}) \right>_{S}} 
\nonumber \\
 & & + \left(\frac{kT}{M}\right)^{2} \frac{1}{N} \sum_{\lambda \lambda'}
\frac{1}{\omega_{\lambda}\omega_{\lambda'}}\left<  h_{\lambda \lambda'}
(\hat{\bm{a}},\bm{q})\;h_{\lambda' \lambda}^{\ast}
(\hat{\bm{a}},\bm{q})\right>_{\bm{q}^{\ast}}
\left<\delta(\omega\pm\omega_{\lambda}\pm\omega_{\lambda'}) \right>_{A},
\end{eqnarray}
where
\begin{equation}\label{eq40}
h_{\lambda\lambda'}(\hat{\bm{a}},\bm{q}) = \sum_{K} e^{-i \bm{q}\cdot \bm{R}_{K}}
e^{-W_{K}(\bm{q})}\;\hat{\bm{a}}\cdot\bm{w}_{K\lambda}\;\bm{q}\cdot
\bm{w}_{K\lambda'},
\end{equation}
and where $\left<\delta(\omega\pm\omega_{\lambda}
\pm\omega_{\lambda'})\right>_{A}$ is the antisymmetric combination 
of four $\delta$-functions,
\begin{eqnarray}\label{eq41}
\left<\delta(\omega\pm\omega_{\lambda}\pm\omega_{\lambda'})\right>_{A}  &= & 
 \frac{1}{4} \Big[\delta(\omega+\omega_{\lambda}+\omega_{\lambda'}) + 
\delta(\omega-\omega_{\lambda}-\omega_{\lambda'})  \nonumber\\
& &  - \delta(\omega+\omega_{\lambda}-\omega_{\lambda'})-
\delta(\omega-\omega_{\lambda}+\omega_{\lambda'})\Big].
\end{eqnarray}
The first line of Eq.~(\ref{eq39}) is the velocity-decoupled term, and
contains $\left<\left| \sum
_{K}\dots\right|^{2}\right>_{\bm{q}^{\ast}}$, similar to
$S^{(2)}(q,\omega)$ in Eqs.~(\ref{eq24}) and (\ref{eq25}). The second
line of Eq.~(\ref{eq39}) is the velocity-interference term, and does
not contain an absolute square, because of the interchange of
$\lambda$ and $\lambda'$ in $h_{\lambda \lambda'}^{\ast}$. It is still
possible to show that the velocity-interference term is overall real,
even without the $\bm{q}^{\ast}$ average. A more prominent difference
from the velocity-decoupled term is the change in sign of two of the
$\delta$-functions in the velocity-interference term. This will give
the two terms a strongly different $\omega$-dependence, since
$\delta(\omega\pm(\omega_{\lambda}+\omega_{\lambda'}))$ is spread
broadly over $\omega$, while $\delta(\omega\pm(\omega_{\lambda}-
\omega_{\lambda'}))$ is concentrated at smaller frequencies.

\section{Summary}
\label{summ}

In V-T theory of monatomic liquid dynamics, the vibrational
contribution to a statistical average can be evaluated exactly from
the vibrational eigenvalues and eigenvectors of a single random
valley. The transit contribution consists of small but not negligible
corrections to the vibrational contribution, and has so far been
accounted for by parametrized models. In the thermodynamic properties
of the elemental liquids, the transit correction accounts for the
constant density entropy of melting \cite{VT5,VT3} and for the
contribution to thermal energy from valley-valley intersections
\cite{VT6}. In nonequilibrium properties, as expressed by time
correlation functions, the transits cause decay of correlations in
addition to that already present in the vibrational contribution
\cite{VT11,ARXIV05,DWJCP123}. Continuing in this line of theoretical
development, we have two goals in the present paper: (a) to derive the
equations for the vibrational contribution to the current
autocorrelation functions, and (b) to address difficulties present in
the liquid state in computing these functions and the related density
autocorrelation functions.

The current autocorrelation functions are denoted
$D(\hat{\bm{a}},q,t)$, where $\hat{\bm{a}}=\hat{\bm{q}}$ for the
longitudinal current, and $\hat{\bm{a}}$ is transverse to
$\hat{\bm{q}}$ for the transverse current. A closed form expression
for $D(\hat{\bm{a}},q,t)$ is derived in Eq.~(\ref{eq34}). This
expression contains two classes of terms: (a) velocity-decoupled
terms, containing one velocity-velocity correlation function, and (b)
velocity-interference terms, containing a product of two
velocity-displacement correlation functions. The one- and two-mode
contributions to the Fourier transform $D(\hat{\bm{a}},q,\omega)$ are
written in Eqs.~(\ref{eq37}-\ref{eq41}):
$D^{(1)}(\hat{\bm{a}},q,\omega)$ expresses decay of current
fluctuations through processes involving one vibrational mode, and
$D^{(2)}(\hat{\bm{a}},q,\omega)$ expresses the same decay through
processes involving two vibrational modes (including one mode twice).

The density autocorrelation function $F(q,t)$, and its Fourier
transform $S(q,\omega)$, are the basis of inelastic neutron scattering
theory for condensed matter. With notable exceptions, the point of
view in crystal theory \cite{MFPR128,ACBLD,CAP12,Lovbook}, and in
amorphous solid theory \cite{CPPRB12}, is that the multimode
expansions of these functions converge well. The exceptions include
the quantum crystals \cite{Glybook}, and soft phonons associated with
the hcp-bcc transition in Ti \cite{SAHWPRB19,PHTAHSVPRB43} and Zr
\cite{SZAMHPRB18,HPTAHSVPRB43}. However, this viewpoint is precarious
for the liquid, because the characteristic liquid state persists to
temperatures of several times $T_{m}$ \cite{VT15}, while the multimode
expansions in question diverge in $T$ at any finite order. The same
convergence issue is present in the current correlations, and the
following conclusions apply to density and current correlation
functions alike.

(a) The multimode series for the vibrational contribution to the
Fourier transforms $S(q,\omega)$ and $D(\hat{\bm{a}},q,\omega)$ are
very informative, as they resolve the decay of fluctuations into
one-mode processes, two-mode processes, and so on. In classical
statistical mechanics, the one-mode processes go as $T$ and are
computable as a double $N$-fold sum; the two-mode processes go as
$T^{2}$ and are computable as a triple $N$-fold sum, and so on. In
general we expect the series to converge at about the same rate for
liquid and crystal phases at $T_{m}$. Then, as $T$ increases above
$T_{m}$, higher order terms which are more difficult to compute become
more important.

(b) These convergence and computability problems are circumvented by
computing the vibrational contributions to the time correlation
functions, $F(q,t)$ and $D(\hat{\bm{a}},q,t)$, and by numerically
transforming these. (We suggest subtracting off $F(q,\infty)$ and
transforming that term analytically, which is not necessary for
$D(\hat{\bm{a}},q,t)$ because $D(\hat{\bm{a}},q,\infty)=0$.)  The real
exponential function in Eq.~(\ref{eq17}) for $F(q,t)$ appears also in
Eq.~(\ref{eq34}) for $D(\hat{\bm{a}},q,t)$, and this function is well
behaved at all $T$.  Eqs.~(\ref{eq17}) and (\ref{eq34}) for the time
correlation functions are computable as triple $N$-fold sums.

\begin{acknowledgments}
We would like to thank Renzo Vallauri for suggesting that it would be useful
to study the current correlation functions in terms of V-T theory.  This work 
was supported by the US DOE through contract DE-AC52-06NA25396.
\end{acknowledgments}

\bibliography{genrefcorr1}

\end{document}